\documentclass[twocolumn,showpacs,preprintnumbers,amsmath,amssymb]{revtex4}
\usepackage{graphicx}
\usepackage{bm}
\usepackage{latexsym}
\usepackage{epsfig}
\usepackage[colorlinks]{hyperref} 
\begin{document}

\title{From de Sitter to de Sitter: decaying vacuum models as a possible solution to the main cosmological problems}

\author{G. J. M. Zilioti}\email{george.zilioti@ufabc.edu.br}
\affiliation{Universidade Federal do ABC (UFABC), Santo Andr\'e, 09210-580, S\~ao Paulo, Brasil}

\author{R. C. Santos}\email{cliviars@gmail.com}
\affiliation{Departamento de F\'isica, Universidade Federal de S\~ao Paulo (UNIFESP) \\ 09972-270  Diadema, SP, Brasil} 

\author{J. A. S. Lima}\email{jas.lima@iag.usp.br}
\affiliation{Departamento de Astronomia, Universidade de S\~ao Paulo (IAGUSP), Rua do Mat\~ao 1226, 05508-900, S\~ao Paulo, Brazil}

\date{\today}

\begin{abstract}
\baselineskip0.4cm
\noindent Decaying vacuum cosmological models evolving smoothly between two extreme (very early and late time) de Sitter phases are capable to solve or at least to alleviate some cosmological puzzles, among them: (i) the singularity, (ii) horizon,  (iii) graceful-exit from inflation, and (iv) the baryogenesis problem.  Our basic aim here is to  discuss how the coincidence problem based on a large class of running vacuum cosmologies evolving from de Sitter to de Sitter can also be mollified. It is also  argued that even the cosmological constant problem become less severe provided that the characteristic scales of the two limiting de Sitter manifolds are predicted from first principles.    

\end{abstract}

\maketitle

\section{Introduction}

The present astronomical observations are being successfully explained by the so-called cosmic concordance model or $\Lambda_0$CDM cosmology \cite{Planck}. However, such a scenario  can hardly  provide by itself a definite explanation for the complete  cosmic evolution involving two unconnected accelerating inflationary regimes separated by many aeons. Unsolved mysteries include the predicted existence of a spacetime singularity in the very beginning of the Universe,  the ``graceful-exit" from primordial inflation, the baryogenesis problem, that is, the matter-antimatter asymmetry and the cosmic coincidence problem. Last but not the least, the scenario is also plagued with the so-called cosmological constant problem \cite{SW89}.  

One possibility to solve such evolutionary puzzles is to incorporate energy transfer among the cosmic components, as happens in decaying or running vacuum models  or, more generally, in the interacting dark energy cosmologies. Here we are interested in the first class of models because the idea of a time-varying vacuum energy density or $\Lambda(t)$-models ($\rho_{\Lambda}\equiv\Lambda(t)/8\pi\,G$) in the expanding Universe is physically more plausible  than the current view of a strict constant $\Lambda$\cite{Bronstein33,OT1986,OT1987,KF87,Carvalho1992,Waga1993,Lima1994,L1996,thesis,Urb1,Urb2}.      

The cosmic concordance model suggests strongly that we live in a flat, accelerating Universe composed of  $\sim 1/3$ of matter (baryons + dark matter) and $\sim 2/3$ of a constant vacuum energy density. The current accelerating period (${\ddot a > 0}$), started at a redshift $z_a \sim 0.69$  or equivalently, when $2\rho_{\Lambda}=\rho_m$. Thus, it is remarkable that the constant vacuum and the time varying matter energy density  are of the same order of magnitude just by now thereby suggesting that we live in a very special moment of the cosmic history. This puzzle (``why now"?) has been dubbed the cosmic coincidence problem (CCP) because the present ratio $\Omega_m/\Omega_{\Lambda}\sim {\mathcal{O}(1)}$,  but it was almost infinite at early times \cite{Stein99,CCPO}. There are many attempts in the literature to solve such a mystery, some of them closely related with interacting dark energy models \cite{CCP1,CCP2,CCP3}. 

Recently, a large class of flat nonsingular FRW type cosmologies, where the vacuum energy density evolves like a truncated power-series in the Hubble parameter $H$, has been discussed in the literature \cite{LBS2013,Perico2013,LBS2015a,LBS2016} (its dominant term behaves like $\rho_{{\Lambda}}(H) \propto H^{n+2}, n > 0$). Such models has some interesting features, among them: (i) a new mechanism for inflation with no ``graceful-exit" problem, (ii) the late-time expansion history is very close to the cosmic concordance model, and (iii) a smooth link between the initial and final de Sitter stages through the radiation and matter dominated phases. 


In this article we will show in detail how the coincidence problem is also alleviated in the context of this class of decaying vacuum models. {In addition, partially based on previous works, we  also advocate here  that a generic running vacuum cosmology providing a complete cosmic history evolving between two extreme de Sitter phases are potentially capable to mitigate several cosmological problems.}

\section{The model: basic equations}
The Einstein equations, $G^{\mu\nu} = 8\pi G\,$ [$T^{\mu\nu}_{(\Lambda)} + T^{\mu\nu}_{(T)}$],  for an interacting vacuum-matter mixture in the FRW geometry read \cite{LBS2013,Perico2013}: 
\begin{eqnarray}
	{8\pi} G\,\rho_T\,+\,\Lambda(H) &=& 3H^{2}\;,\label{eq5}\\
	8\pi G\, p_T - \Lambda(H) &=& -2\dot H - 3H^{2}\,,\label{eq6}
	\end{eqnarray}
where $\rho_T=\rho_M + \rho_R$ and $p=p_M + p_R$ are the total energy density and pressure of the material medium formed by nonrelativistic matter and radiation. Note that the bare $\Lambda$ appearing in the geometric side was absorbed on the matter-energy side in order to describe the effective vacuum  with energy density $\rho_{\Lambda} = -p_{\Lambda} \equiv \Lambda(H)/8\pi G$. Naturally, the time dependence of $\Lambda$ is provoked by the vacuum energy transfer to the fluid component. In this context, the total energy conservation law, $u_{\mu}[T^{\mu\nu}_{(\Lambda)} + T^{\mu\nu}_{(T)}]_{;\nu}=0$,  assumes the following form: 
\begin{equation}
\dot \rho_T + 3H (\rho_T + p_T) = -{\dot \rho_{\Lambda}} \equiv -{\frac{\dot \Lambda}{8 \pi G}}.
\end{equation}
{\it What about the behavior of $\dot \Lambda$?} Assuming that the created particles have zero chemical potential and that the vacuum fluid behaves like a condensate carrying no entropy,  as happens in the Landau-Tisza two-fluid description employed in helium superfluid dynamics\cite{LL85}, it has been shown that $\dot \Lambda < 0$ as a consequence of the second law of thermodynamics \cite{L1996}, that is, the vacuum energy density diminishes in the course of the evolution.  Therefore, in what follows we consider that the coupled vacuum  is continuously transferring energy to the dominant component (radiation or nonrelativistic matter components). Such a property defines precisely the physical meaning of decaying or running vacuum cosmologies in this work.   
 
Now, by combining the above field equation it is readily checked that:

\begin{equation}
 \dot H + \frac{3(1+\omega)}{2}H^{2} - \frac{1+\omega}{2} \Lambda(H)=0\,,
\end{equation}
where the equation of state $p_T=\omega \rho_T$ ($\omega \geq 0$) was used.  The above equations are solvable only if we know the functional form of $\Lambda(H)$. 

The decaying vacuum law adopted here was first  proposed based on phenomenological grounds \cite{Carvalho1992,Waga1993,Lima1994,thesis}, and latter on  suggested by the renormalization group approach techniques applied to quantum field theories in curved spacetimes \cite{Shapiro2002}. It is given by:  

\begin{equation}\label{PL}
\Lambda(H) \equiv 8\pi G\rho_{\Lambda} = c_0 + 3\nu H^{2} + \alpha \frac{H^{n+2}}{{H_I}^{n}}\,,
\end{equation}   
where $H_I$ is an arbitrary time scale describing the primordial  de Sitter era (the upper limit of the Hubble parameter),  $\nu$ and $\alpha$  are dimensionless constants,  and $c_0$ is a constant  with dimension of $[H]^{2}$.  

In a point of fact, the constant $\alpha$ above does not represent a new degree of freedom. It can be determined with the proviso that for large values of $H$, the model starts from a de Sitter phase  with $\rho=0$  and $\Lambda_I = 3H^{2}_I$. In this case, from (\ref{PL}) one finds  $\alpha = 3(1-\nu)$ because the first two terms there are negligible in this limit [see  Eq. (1) in \cite{Lima1994} for the case $n=1$ and \cite{thesis} for a general $n$]. The constant $c_0$ can be fixed by the time scale of the final de Sitter phase.   For $H << H_I$  we also see from (4) that $c_0= 3(1-\nu)H^{2}_F$, where $H_F$  characterizes the  final de Sitter stage (see Eqs. (6) and (8)).  Hence, the phenomenological law (\ref{PL}) assumes the final form:

\begin{equation}\label{Lf}
\Lambda(H) = 3(1-\nu)H^{2}_F + 3\nu H^{2} + 3(1-\nu) \frac{H^{n+2}}{{H_I}^{n}}.
\end{equation} 
This is an interesting 3-parametric phenomenological expression.  It depends on the arbitrary dimensionless constant $\nu$ and also of the two extreme Hubble parameters  ($H_I$, $H_F$) describing the primordial and late time inflationary phases, respectively. Current observations imply that the value of $\nu$ is very small, $|\nu| \sim 10^{-6} - 10^{-3}$ \cite{AS2012,AS2015}. More interesting, the analytical results discussed below remain valid even for $\nu=0$.  In this case, we obtain a sort of minimal model defined only by a pair of physical time scales, $H_I$ and $H_F$, determining the entire evolution of the Universe.  As we shall see, the possible existence of these two extreme de Sitter regimes suggest a different  perspective  to the cosmological constant problem. 
  
By inserting the above expression into (3) we obtain the equation of motion:

\begin{equation}\label{motion}
\dot H + \frac{3(1+\omega)(1-\nu)}{2}H^{2}\left[1 - \frac{H^{2}_F}{H^2} - \frac{H^{n}}{H^{n}_I}\right]=0.
\end{equation}
In principle,  all possible de Sitter phases here are simply characterized by a constant Hubble parameter ($H_C$) satisfying the conditions $\dot H =\rho = p = 0$ and $\Lambda = 3H^{2}_{C}$. For all physically relevant values of  $\nu$ and $\omega$ in the present context, we see that the condition $\dot H=0$ is satisfied whether the possible values of $H_C$ are constrained by the algebraic equation involving  the arbitrary  (initial and final) de Sitter vacuum scales $H_I$ and $H_F$: 

\begin{equation}\label{algeb}
H^{n+2}_C - H^{n}_{I} H^{2}_C + H^{2}_{F} H^{n}_{I} = 0\,.
\end{equation}
In particular, for $n=2$, the value preferred from the covariance of the action, the exact solution is given by:
\begin{equation}
 H^{2}_C = \frac {H^{2}_I}{2}  \pm  \frac{H^2_{I}}{2} \sqrt{1 - 4H^{2}_{F}/H^{2}_I}\,, 
\end{equation}
and since $H_F << H_I$ we see that the two extreme scaling solutions for $n=2$ are $H_{1C}=H_I$ and $H_{2C}=H_F$. However, we also see directly from (\ref{algeb}) that  the condition   $H_F << H_I$,  also guarantees that such solutions are valid regardless of the values of $n$.  In certain sense,  since $H_0$ is only the present day expansion rate, characterizing a quite casual stage of the recent evolving Universe, probably, it is not the interesting scale to be a priori predicted. In what follows we consider that the pair of extreme de Sitter scales ($H_I, H_F$) are the physically relevant quantities. This occurs because different from $H_0$, the expanding de Sitter rates are associated with very specific limiting manifolds. For instance, it is widely known that de Sitter spaces are static when written in a suitable coordinate system. {Besides the discussion on the coincidence problem (see next section),  a new idea to be advocated here} is that the prediction of such scales, at least in principle, should be an interesting theoretical target. Their first principles prediction would open a new and interesting route to investigate the cosmological constant problem.   

The solutions for the Hubble parameter describing analytically the transitions vacuum-radiation ($\omega = 1/3$) and matter-vacuum ($\omega=0$) can be expressed in terms of the scale factor, the couple of scales ($H_I, H_F$) and free parameters ($\nu, n$):  

\begin{eqnarray} \label{solutions}
H &=& \frac{H_I}{[1 + Ca^{2n(1-\nu)}]^{1/n}}\,,\\
H &=& {H_F}[Da^{-3(1-\nu)} + 1]^{1/2}\,. 
\end{eqnarray}
We remark that the transition radiation-matter is like in the standard  cosmic concordance model.  The only difference is due to the small $\nu$ parameter that can fixed to be zero (minimal model). Indeed, if one fixes $\nu =0$, the matter-vacuum transition is exactly  the same one appearing in the flat $\Lambda$CDM model. As we shall see below, the final scale $H_F$ can be expressed as a simple function of $H_0$, $\nu$ and $\Omega_{\Lambda}$. Naturally, the existence of such an expression is needed in order to compare with the present observations. However, it cannot be used to hide the special meaning  played by $H_F$ in a possible solution of the cosmological constant problem. 

\section{Alleviating the coincidence problem}

The so-called coincidence problem is very well known.  It comes from the fact that the matter energy density of the nonrelativistic components   (baryons + dark matter) decrease as the universe expands while the vacuum energy density ($\rho_{\Lambda_0}$) is always constant in the cosmic concordance model ($\Lambda_0$CDM). This happens also because the energy density of the radiation $\rho_{\gamma}$ (CMB photons) and neutrinos ($\rho_{\nu}$) are negligible today. Thus, in a broader perspective, one may also say that the ratio $(\rho_M + \rho_R) /\rho_{\Lambda_0}$, where $\rho_R = \rho_{\gamma} + \rho_{\nu}$, was almost infinite at early times, but it is nearly of the order unity today.

The current fine-tuning behind  the coincidence problem   can also be readily defined in terms of the  corresponding density parameters, since  ($\Omega_{\Lambda_0} \sim 0.7$ and  $\Omega_M + \Omega_R \sim 0.3$), so that the ratio is of the order unity some 14 billion years later. 

\begin{figure}[th]
  \begin{minipage}[b]{0.45\textwidth}
    \includegraphics[width=\textwidth]{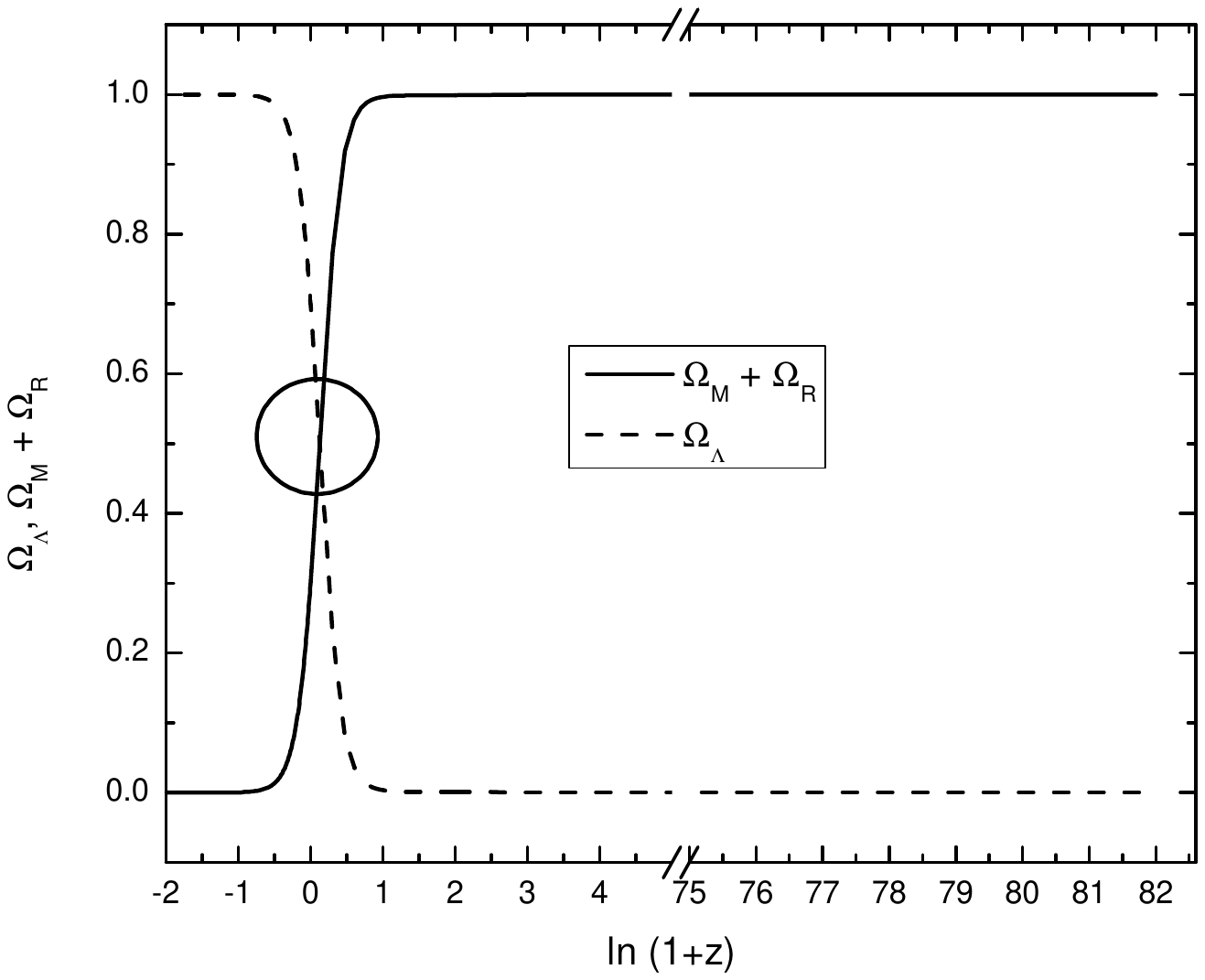}
  \end{minipage}
  \caption{Standard coincidence problem in the cosmic concordance model ($\Lambda_0$CDM). Solid and dashed lines represent the evolution of the vacuum ($\Omega_{\Lambda_0}$) and  total matter-radiation  ($\Omega_M + \Omega_R$) density parameters.  The circle marks the  low (and unique!) redshift presenting the extreme coincidence between the density parameter of the vacuum and material medium. Note that the model discussed here is fully equivalent to $\Lambda_0$CDM when the time dependent corrections in the decay $\Lambda(t)$ expression  are neglected [see equation (5)].}  
\end{figure}
{\vskip 0.3cm}


In {Figure 1} we display the standard view of the coincidence problem in terms of the corresponding density parameters: $\Omega_M = \Omega_b + \Omega_{cdm}$ (baryons + cold dark matter) and $\Omega_R = \Omega_{\gamma} + \Omega_{\nu}$ (CMB photons + relic neutrinos). 
As one may conclude from the figure, the ratio  was practically infinite at very high redshifts, that is, at the early Universe (say, roughly at the Planck time). However, both densities are nearly coincident at present. The ratio $(\Omega_{R}+\Omega_{M})/\Omega_{\Lambda_0}  \sim 1$) at low redshifts. Note also that in the far future, that is, very deep in the de Sitter stage, the ratio approaches zero or equivalently,  the inverse ratio is almost infinite  because the vacuum component becomes fully dominant.


A natural  way to solve this puzzle is to assume  that the vacuum energy density must vary in the course of the expansion. As shown in the previous section, 
the characteristic scales of the $\Lambda(t)$ model specify the evolution during the extreme de Sitter phases: the primordial vacuum solution with $Ca^{2n(1-\nu)} << 1$  and $H=H_I$,  behaves like a ``repeller" in the distant past, while the final vacuum solution for $a>>1$, that is,  $Da^{3(1-\nu)} \mapsto 0$ and $H=H_F$ is an attractor in  the distant future. 

The arbitrary integration constants C and D are also easily determined.  The constant C can be fixed by the end of the primordial inflation ($\ddot a=0$) or equivalently $\rho_{\Lambda}=\rho_R$. This means that   $C=a_{(eq)}^{-2n(1-\nu)}/(1-2\nu)$  [$a_{(eq)}$ corresponds to the value of the scale factor at vacuum-radiation equality]. In terms of the present day observable quantities we also find $D =\Omega_{M0}/(\Omega_{{\Lambda}0} - \nu)$ and $H_{F} = H_0\sqrt{\Omega_{{\Lambda}0} - \nu}/\sqrt{1 -\nu}$.  For $\nu=0$ and $\Omega_{{\Lambda}0} \sim 0.7$ one finds $H_F \sim 0.83H_0$, as expected a little smaller than $H_0$. The small observable parameter $\nu < 10^{-3}$ quantifies the difference between the late time decaying vacuum model and the cosmic concordance cosmology, namely:
\begin{equation}
H = \frac{H_0}{\sqrt{1-\nu}}[\Omega_{M0}a^{-3(1-\nu)} + 1-\Omega_{M0}-\nu]^{1/2}\,.
\end{equation}
As remarked above,  the $H(a)$ expression of the standard $\Lambda$CDM model  is fully recovered for $\nu=0$.  

The solution of the coincidence problem in the present framework can be  demonstrated as follows. The density parameters of the vacuum and material medium are given by:  

\begin{eqnarray}
\Omega_{\Lambda} \equiv \frac{\Lambda(H)}{3H^{2}} = \nu + (1-\nu)\frac{H^{2}_F}{H^{2}} + (1-\nu)\frac{H^{n}}{H^{n}_I}\,,\\
\Omega_T \equiv 1- \Omega_{\Lambda} = 1-\nu  - (1-\nu)\frac{H^{2}_F}{H^{2}} - (1-\nu)\frac{H^{n}}{H^{n}_I}\,.
\end{eqnarray}

Such results are a simple consequence of  expression (\ref{Lf}) for $\Lambda(H)$ and the constraint Friedman equation (1). Note that $\Omega_T \equiv \Omega_M + \Omega_R$ is always describing the dominant component, either the nonrelativistic matter ($\omega=0$) or radiation ($\omega=1/3$).

The density parameters of the vacuum and material medium are equal in two different epochs specifying the dynamic transition between the distinct dominant components. These specific moments of time  will be  characterized here by Hubble parameters $H^{eq}_1$ and $H^{eq}_2$.  The first equality (vacuum-radiation, $\rho_{\Lambda}=\rho_R$) occurs just  at the end of the first accelerating stage ($\ddot a = 0$), that is,  when $H^{eq}_1 = [\frac{1-2\nu}{2(1-\nu)}]^{1/n} H_I$, while the second one is at low redshifts when  $H^{eq}_{2} = [\frac{2(1-\nu)}{1-2\nu}]^{1/2} H_F$. Note that such results are also valid for the minimal model by taking  $\nu=0$. In particular, inserting $\nu=0$ in the first expression above  we find $H^{eq}_1 = H_I/2^{1/n}$. The scale $H^{eq}_2$ can also be determined in terms of $H_0$. By adding the result $H_F \sim 0.83 H_0$ we find for $\nu=0$ that $H^{eq}_2 \sim 1.18H_0$,  which is higher than $H_0$, as should be expected for the matter-vacuum transition.   

\begin{figure}[th]
  \begin{minipage}[b]{0.45\textwidth}
    \includegraphics[width=\textwidth]{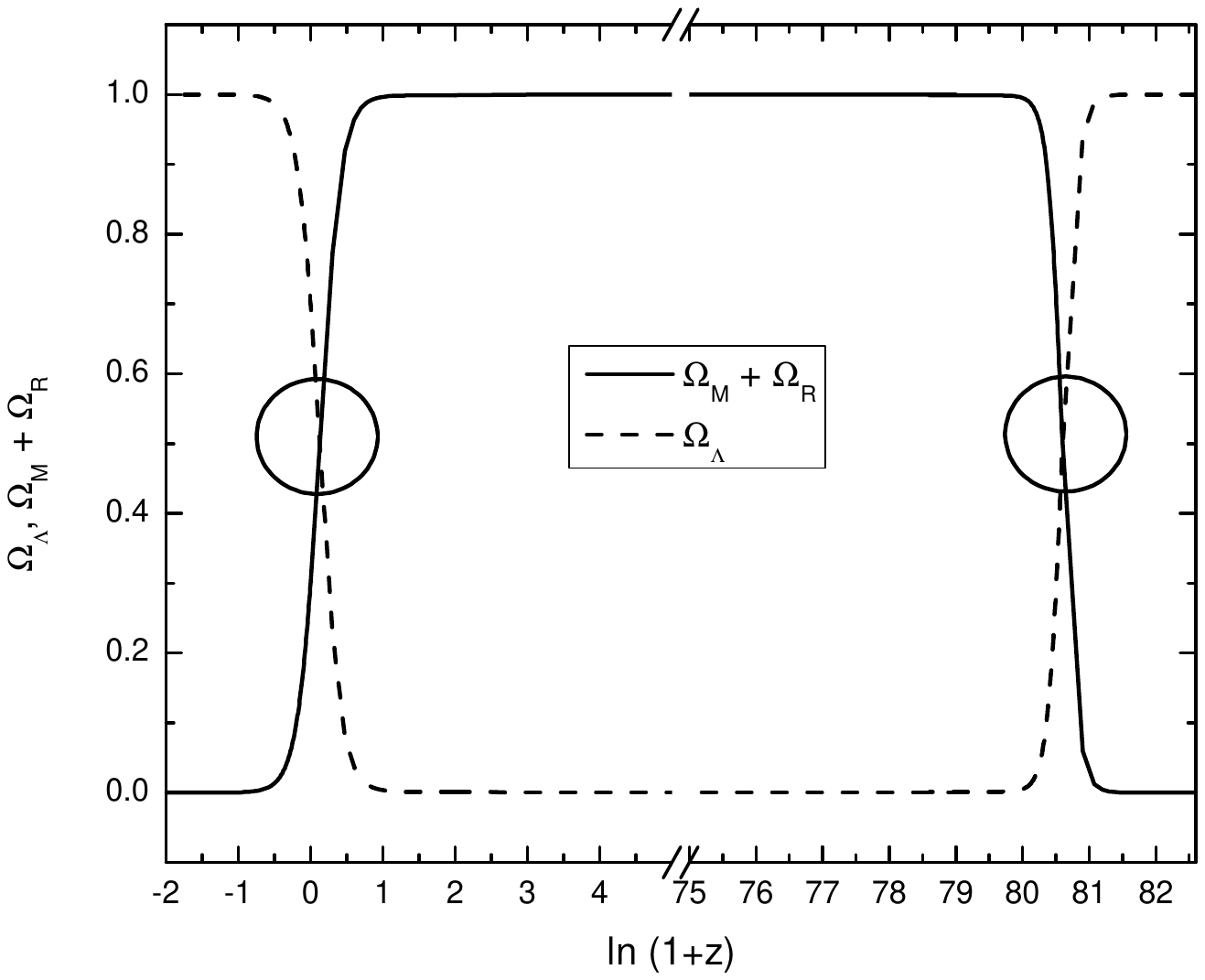}
  \end{minipage}
\caption{Solution of the coincidence problem in running vacuum cosmologies. The right graphic is our model, the left is $\Lambda$CDM. Solid and dashed lines represent the evolution of the vacuum ($\Omega_{\Lambda}$) and  total matter-radiation  ($\Omega_M + \Omega_R$) density parameters for n=2, $\nu = 10^{-3}$ and $H_I/H_0 =10^{60}$. The late time coincidence  between the density parameter of the vacuum and material medium (left circle)  has already occurred at very early times (right circle). Note also that the values  5 and 75 in the horizontal axis were glued in order to show the complete evolution (the suppressed part present exactly the same behavior).  Different values of $n$ changes slightly the value of the redshift for which $\Omega_{\Lambda} = \Omega_M + \Omega_R$ at the very early Universe (see also discussion in the text).} 
\end{figure}

Naturally,  the existence of two subsequent equalities on the density parameter  suggests a solution to the coincidence problem.  Neglecting terms of the order of $10^{-120}$ and $10^{-60n}$ in above expressions, it is easy to demonstrate  the following results:
\vspace{0.4cm}

1) $Lim_{H\rightarrow H_I}{\Omega_{\Lambda}}\, = 1$  and  $Lim _{H\rightarrow H_I}{\Omega_{T}}=0$,

2) $Lim_{H\rightarrow H_F}{\Omega_{\Lambda}}\, = 1$  and $Lim _{H\rightarrow H_F}{\Omega_{T}} = 0$.

{\vskip 0.3cm}

\noindent The meaning of the above results is quite clear. The density parameters of the vacuum and material components (radiation + matter)  perform a cycle, that is, $\Omega_{\Lambda}$, and $\Omega_M  + \Omega_R $ are periodic in the long run.  

In { Figure 2}, we show the complete evolution of the vacuum and matter-energy density parameters for this class of decaying vacuum model. Different from { Figure 1} we observe that the values of  $\Omega_{\Lambda}$ and $\Omega_M + \Omega_R$ are  cyclic in the long run.

These parameters  start and finish the evolution satisfying the above limits. The physical meaning of such evolution is also remarkable. For any value of $n>0$, the model starts as a pure unstable vacuum  de Sitter phase with $H=H_I$ (in the begin there is no matter or radiation, $\Omega_{\Lambda} = 1, \Omega_M + \Omega_R = 0$). The vacuum decays  and the model evolves smoothly to a quasi-radiation phase parametrized by the small $\nu$-parameter. 

The circles show the redshifts for which $\Omega_{\Lambda} = \Omega_M + \Omega_R$. Of course, the existence of two equality solutions alleviates the cosmic coincidence problem. 

The robustness of the solution must also be commented. It holds not only for any value of $n > 0$ but also for $\nu=0$. In the later case,  the primordial nonsingular vacuum state deflates directly to the standard FRW  radiation phase. Later on, also occurs the transition from radiation to matter-vacuum dominated phase, thereby reproducing exactly the matter-vacuum transition of the standard $\Lambda_0$CDM model.  

The ``irreversible entropic cycle" from  initial Sitter ($H_I$) to the late time  de Sitter stage  is completed when the Hubble parameter approaches its small final value ($H \mapsto H_F$).  The de Sitter spacetime that was a `repeller' (unstable solution) at very early times ($z \rightarrow \infty$)  becomes an attractor in the distant future ($z  \rightarrow -1$) driven by the incredibly low energy scale $H_F$ which is associated to the late time vacuum energy density, $\rho_M \rightarrow 0, \,\rho_{\Lambda F} \propto H_F^{2}$. 

Like the above solution to the coincidence problem, some cosmological puzzles can also be resolved along  the same lines because the time behavior of the present scenario  even fixing $\alpha = 1- \nu$ has been proven here to be  exactly the one discussed in Ref. \cite{Perico2013} (see also \cite{Lima1994} for the case $n=1$). 

\section{Final Comments and Conclusion}

As we have seen, the phenomenological $\Lambda(t)$-term provided a possible solution to the coincidence problem because the ratio $\Omega_M/\Omega_{\Lambda}$ is periodic in long run (see {Figure 2}). In other words, the coincidence  is not a novelty exclusive of the current epoch (low redshifts) since it also happened  in the very early Universe at extremely high redshifts. In this framework,  such a result seems to be robust because it is not altered even to the minimal model, that is, for $\nu=0$.

{It should also be stressed that the alternative complete cosmological scenario (from de Sitter to de Sitter) is not a singular attribute of decaying  vacuum models. For instance, it was recently proved that at the background level such models are equivalent to gravitationally induced particle production cosmologies \cite{Prigogine89, CLW92} by identifying ${\Lambda(t)}\equiv \rho\Gamma/3H$, where $\Gamma$ is the gravitational particle production rate.}  In a series of papers \cite{LJO2010,LBC2012}, the dynamical equivalence of such scenario at late times with the cosmic concordance model  was also discussed. {It is also interesting that such a reduction of the dark sector  can mimic the cosmic concordance model ($\Lambda_0$CDM)  both at the background and perturbative levels \cite{RSW2014a,RSW2014b}.  In principle, this means that alternative scenarios evolving  smoothly between two extreme de Sitter phases are also potentially able  to provide viable solutions of the main cosmological puzzles. {However, different from $\Lambda(t)$-cosmologies, such alternatives are unable to explain the cosmological constant problem  with  this extreme puzzle becoming restricted to the realm of quantum field theory.} 

At this point, in order to compare our results  with alternative models also evolving between two extreme de Sitter stages, it is interesting to review briefly how the main cosmological problems are solved (or alleviated) within this class of models driven by a pure decaying vacuum initial state:
\begin{itemize}

\item {\it Singularity:} The spacetime in the distant past is a nonsingular de Sitter geometry with an arbitrary energy scale $H_I$.  In order to agree with the semi-classical description of gravity, the arbitrary scale $H_I$ must be constrained by the upper limit  $H_I \leq 10^{19}$ GeV (Planck energy) in natural units, or equivalently,  the the based on general relativity is valid only for times greater than the Plank time, $H^{-1}_I \geq  10^{-43}$ sec. 

\item {\it Horizon Problem:} The ansatz  (\ref{Lf}) can mathematically be considered as the
simplest decaying vacuum law which destabilizes the initial de Sitter configuration. Actually, in such a model the Universe begins  as a steady-state cosmology, $R \sim e^{H_It}$. Since the model is nonsingular, it is easy to show that the horizon problem is naturally solved in this context (see, for instance, Refs. \cite{LBS2016}).

\item {\it ``Graceful-Exit" from Inflation:} The transition from the early de Sitter to the radiation phase is smooth and driven by Eq. (10). The first coincidence of density parameters happens for  $H = H^{eq}_{1},\, \rho_{\Lambda} = \rho_R$ and $\ddot a=0$, that is, when the first inflationary period ends (see {Figure 2}).  All the radiation entropy ($S_0 \sim 10^{88}$, in dimensionless units) and matter-radiation content now observed were generated during the early decaying vacuum  process (see \cite{LBS2015a} for the entropy produced in the case $n=2$). For an arbitrary $n>0$, the exit of inflation and the entropy production  has also already been discussed \cite{LBS2016}. Some possible curvature  effects were also analyzed \cite{LPZ}.

\item{\it Baryogenesis problem:} Recently, it was shown that the matter-antimatter asymmetry can also be induced by a derivative coupling between the running vacuum and a non-conserving baryon current \cite{LS2017,LS2018}. Such an ingredient breaks dynamically CPT thereby triggering baryogenesis through an effective chemical potential  (for a different but related approach see \cite{OSP2017}).  Naturally, baryogenesis induced by a running vacuum process has at least two interesting features:   (i) The variable vacuum energy density is the same ingredient driving the early accelerating phase of the Universe and it also control the baryogenesis process; (ii) the running vacuum is always accompanied by particle production and entropy generation \cite{Waga1993,L1996,LBS2016}. This nonisentropic process is an extra source of T-violation (beyond the freeze out of the B-operator) which as first emphasized by Sakharov \cite{Shaka1967} is a basic ingredient for successful baryogenesis. In particular, for $\nu=0$ it was found that the observed B-asymmetry ordinarily quantified by the $\eta$ parameter
 \begin{equation}
\label{eta}
5.7\times 10^{-10} < \eta  < 6.7 \times 10^{-10}\,,
\end{equation}
can be obtained for a large range of the relevant parameters ($H_I, n$) of the present model\cite{LS2017, LS2018}. Thus, as remarked before,  the proposed running vacuum cosmology may also provide a successful baryogenesis mechanism. 

\item {\it de Sitter Instability and the future of the Universe:} Another interesting aspect associated with the presence of two extreme  Sitter phases as discussed here,  are the intrinsic instability of such spacetimes. Long time ago, Hawking showed that the spacetime of a static black hole is thermodynamically unstable to macroscopic fluctuation in the temperature of the horizon \cite{Hawking1976}. Later on, it was also demonstrated by Mottola \cite {Mottola1986} based on the validity of the generalized second law of thermodynamics that the same arguments used by Hawking in the case of black holes remain valid for the de Sitter spacetime. In the case of the primordial de Sitter phase, described here by the characteristic scale $H_I$, such an instability is dynamically described by  solution (10) for $H(a)$. As we know, it behaves like a `repeller' driving the model to the radiation phase. However, the instability result in principle must also be valid to the final de Sitter stage which  behaves like an attractor. In this way, once the final de Sitter phase is reached, the spacetime would evolve to an energy scale smaller than $H_F$ thereby starting a new evolutionary   `cycle' in the long run.  

\item  {\it Cosmological Constant Problem:} It is known  that phenomenological decaying vacuum models are unable to solve this conundrum \cite{LBS2016,LS2017}. The basic reason seems to be related with the clear impossibility to predict the present day value of the vacuum energy density (or equivalently the value of $H_0$) from first principles. However, the present phenomenological approach can provide a new line of inquire in the search for alternative (first principle)  solutions for this remarkable puzzle. In this concern, we notice that the minimal model discussed here depends only on two relevant physical scales ($H_F, H_I$) which are associated to the extreme de Sitter phases. The existence of such scales implies that the ratio between the late and very early  vacuum energy densities   $\rho_{{\Lambda}F}/\rho_{{\Lambda}I} = (H_F/H_I)^{2}$  does not depend explicitly on the Planck mass. Indeed, the gravitational constant (in natural units, $G=M^{-2}_{Planck}$)  arising in the expressions of  the early and late time  vacuum energy densities cancels out in the above ratio. Since $H_F \sim 10^{-42}$GeV, by assuming that $H_I \sim  10^{19}$GeV (the cutoff of classical theory of gravity), one finds that the ratio  ${\rho_{\Lambda F}}/{\rho_{\Lambda I}} \sim 10^{-122}$, as suggested by some estimates based on quantum field theory, {a result already obtained in some nonsingular decaying vacuum models \cite{LBS2013}.} In this context, the open new perspective is  related to the search for a covariant action principle  where both scales arise naturally. One  possibility is related with models whose  theoretical foundations are based on modified gravity theories like $F(R), F(R,T), etc$ [see  for instance, Refs. \cite{Sot2010,Harko2011}].
\end{itemize}

The results outlined above suggest that decaying vacuum models phenomenologically described by $\Lambda(t)$-cosmologies may be considered  an interesting alternative to the mixing scenario formed by the standard $\Lambda$CDM plus inflation. However, although justified from different viewpoints, the main difficulty of such models seems to be a clear-cut covariant Lagrangian description.


\vskip 0.3cm
{\bf Acknowledgments:}  GZ is grateful to a fellowship from CAPES (Brazilian Research Agency), RCS thanks the INCT-A project, and  JASL is partially supported by  CNPq, FAPESP (LLAMA project), and CAPES (PROCAD 2013). The authors are grateful to Spyros Basilakos, Joan Sol\`a  and Douglas Singleton for helpful discussions.\\

The authors declare that there is no conflict of interest regarding the publication of this paper.
\vskip 0.3cm

\end{document}